%% file: main.tex
\documentclass[sigconf, nonacm]{acmart}

\setcopyright{acmcopyright}
\copyrightyear{2024}
\acmYear{2024}
\acmDOI{-}

\acmConference[Conference '24]{}{}{}
\acmBooktitle{}
\acmISBN{978-1-4503-XXXX-X/18/06}

\usepackage{amsmath,amsfonts}
\usepackage{algorithmic}
\usepackage{graphicx}
\usepackage{textcomp}
\usepackage{xcolor}
\usepackage{balance}
\usepackage{xspace}
\usepackage{subcaption}
\usepackage{amsthm}
\usepackage{enumitem}
\usepackage{tabulary}
\usepackage{hyperref}
\usepackage{cleveref}
\usepackage{booktabs}
\usepackage{tikz}
\usetikzlibrary{arrows}
\usetikzlibrary{shapes}

\theoremstyle{definition}
\newtheorem{definition}{Definition}[section]
\newcommand{\method}{{\em OmniMatch}\xspace}

\newcommand{\para}[1]{\vspace{1mm}\noindent\textbf{#1.}}

\newcommand{\parait}[1]{\vspace{.9mm}\noindent\emph{#1.}}

\newcommand{\mybox}[1]
{
\vspace{1.8mm}
\noindent \hspace{-1mm} 
\setlength\fboxsep{1mm}
\fbox{\parbox{\dimexpr\linewidth-2\fboxsep-2\fboxrule}{\itshape #1}}
}

\newcommand{\myBoldTag}[1]{ {\bf #1}}
\def\BibTeX{{\rm B\kern-.05em{\sc i\kern-.025em b}\kern-.08em
    T\kern-.1667em\lower.7ex\hbox{E}\kern-.125emX}}
\usepackage[disable]{todonotes}
\newcommand{\asterios}[1]{\todo[inline, color=red!20]{Asterios: #1}}

\newcommand{\jiani}[1]{\todo[inline, color=blue!20]{Jiani: #1}}
\newcommand{\reviewer}[1]{\todo[inline, color=purple!20]{reviewer #1}}

\begin{document}

\title{OmniMatch: Effective Self-Supervised Any-Join Discovery in Tabular Data Repositories}

\author{Christos Koutras$^{\circ,\triangle}$
 \qquad Jiani Zhang$^\circ$
 \qquad Xiao Qin$^\circ$
 \qquad Chuan Lei$^\circ$
\qquad Vasileios Ioannidis$^\circ$ \\
 Christos Faloutsos$^{\circ,\Box}$
 \qquad George Karypis$^\circ$
 \qquad Asterios Katsifodimos$^{\circ,\triangle}$ \vspace{2.5mm} \\
 \textsuperscript{$\triangle$}\textit{Delft University of Technology} \quad \textsuperscript{$\Box$}\textit{Carnegie Mellon University} \quad \textsuperscript{$\circ$}\textit{Amazon Web Services} \\
c.koutras@tudelft.nl, \{zhajiani,drxqin,chuanlei,ivasilei,faloutso,gkarypis,akatsifo\}@amazon.com 
 }
\renewcommand{\shortauthors}{C. Koutras, J. Zhang, X. Qin, C. Lei, V. Ioannidis, C. Faloutsos, G. Karypis, and A. Katsifodimos}

\begin{abstract}
How can we discover join relationships among columns of tabular data in a data repository? Can this be done effectively when metadata is missing? Traditional column matching works mainly rely on similarity measures based on exact value overlaps, hence missing important semantics or failing to handle noise in the data. At the same time, recent dataset discovery methods focusing on deep table representation learning techniques, do not take into consideration the rich set of column similarity signals found in prior matching and discovery methods. Finally, existing methods heavily depend on user-provided similarity thresholds, hindering their deployability in real-world settings.

In this paper, we propose \method, a novel join discovery technique that detects equi-joins and fuzzy-joins between columns by combining column-pair similarity measures with Graph Neural Networks (GNNs). \method's GNN  can capture column relatedness leveraging graph transitivity, significantly improving the recall of join discovery tasks. At the same time, \method{} also increases the precision by augmenting its training data with negative column join examples through an automated negative example generation process. Most importantly, compared to the state-of-the-art matching and discovery methods, \method{} exhibits up to {\bf 14\%} higher effectiveness in F1 score and AUC without relying on metadata or user-provided thresholds for each similarity metric.
\end{abstract}

\maketitle

\input{1_Introduction}

\input{2_Problem_Definition}
\input{3_Approach_Overview}
\input{4_Graph_Construction}
\input{5_Model_Training}
\input{6_Experimental_Evaluation}
\input{7_Related_Work}
\input{8_Conclusion}

\balance

\bibliographystyle{ACM-Reference-Format}
\bibliography{references}

\balance

\end{document}

%% file: 1_Introduction.tex
\section{Introduction}
\label{sec:introduction}

How can we accurately detect join relationships among columns in a repository of tabular data? Is it possible to identify both equi- and fuzzy-joins in the data?  Can we \textit{effectively} discover such joins even when the quality of the metadata is low, or the metadata is missing?

Organizations are creating and maintaining numerous uncurated \textit{data repositories}, which are rendered less valuable due to the absence of relatedness metadata. These data repositories mainly comprise tabular data, such as relational data from databases, and semi-structured data, including CSV files and spreadsheets. They often contain valuable information for various stakeholders. The column joinability relationships are among the most critical types of relatedness metadata across tabular datasets. 
\emph{Joins} play a vital role in facilitating the exploration and exploitation of datasets. For instance, data scientists, who train machine learning (ML) models on specific datasets, can leverage joins to identify related datasets  that provide additional features, thereby improving the accuracy of an ML model \cite{chepurko13arda, zhao2022leva}. In addition, joins can aid in data cleaning by enabling the discovery of new sources of information that serve as ground truth for error checking, inferring missing values, or eliminating duplicates.

\begin{figure}[t]
    \centering
    \includegraphics[width=\columnwidth]{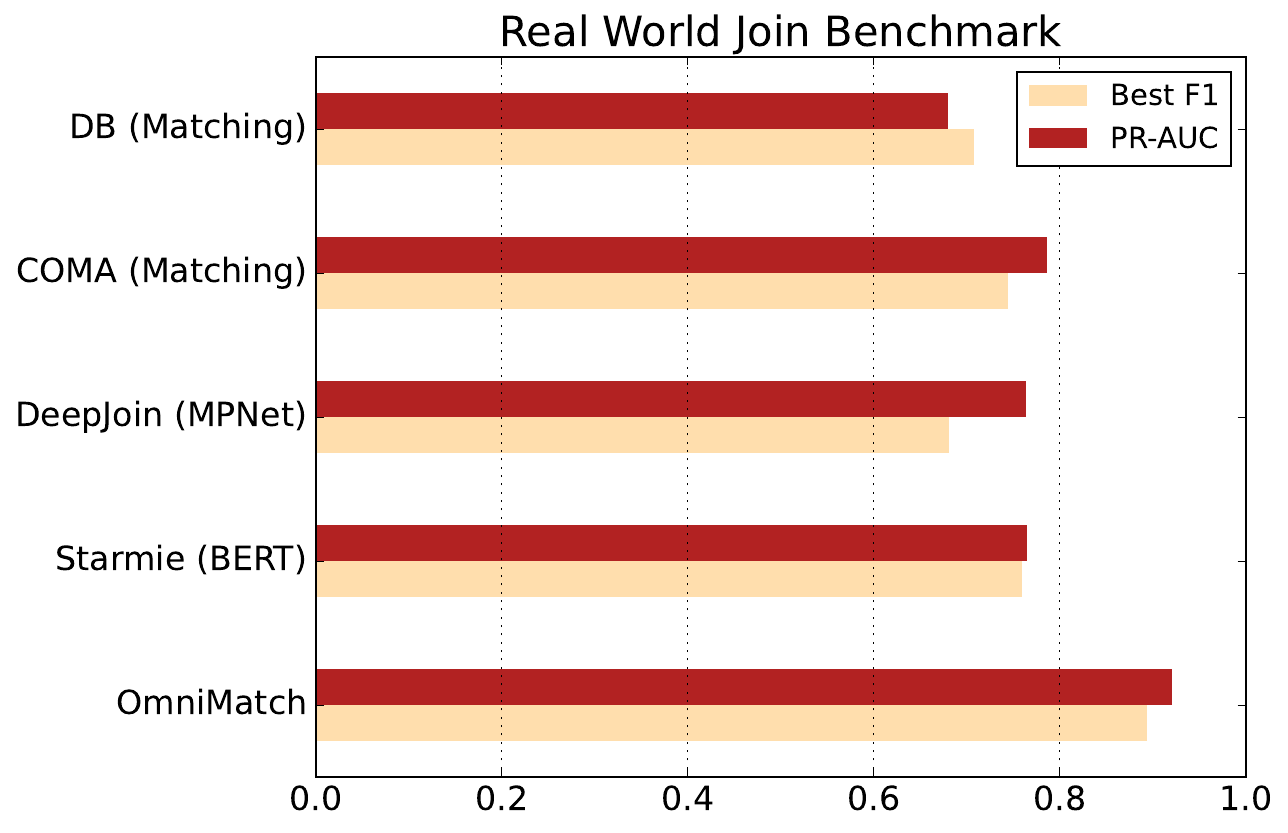}
    \caption{\method{} outperforms the state-of-the-art column matching and representation methods in terms of best F1 and Precision-Recall AUC scores achieved when tested upon real-world join benchmarks on open data repositories (\S~\ref{sec:exp}). Best viewed in color.}
    \label{fig:comp_first}
\end{figure}

\para{Challenges in Join Discovery} A join between two columns entails an overlap among their values, which should also refer to the same domain. However, it is often difficult to quantify value overlaps since using fixed thresholds on set similarity metrics, such as Jaccard Index, might increase false negative/positive rates (due to high/low thresholds respectively). Importantly, joins between columns can exist even when their contents differ syntactically. In the literature, those are termed \emph{fuzzy joins}~\cite{wang2011fast, afrati2012fuzzy, wang2019mf, chen2019customizable, li2021auto}. Fuzzy joins require similarity metrics that capture relatedness beyond value overlaps. This raises the question of which similarity metrics should be used to ensure high effectiveness. Finally, without metadata, such as column names and descriptions about tabular data, finding joins requires understanding the value semantics.

\para{Existing Solutions} Existing join discovery methods primarily fall into the domain of \emph{schema matching} \cite{rahm2001survey}. These methods aim to find column correspondences between tabular datasets using various techniques such as leveraging metadata \cite{madhavan2001generic, melnik2002similarity}, instances \cite{zhang2011automatic} or both \cite{do2002coma}. Notably, language models have also been utilized to create column representations for finding column matches by using pre-trained word-embeddings \cite{fernandez2018seeping}, training skip-gram models on the domain-specific datasets \cite{cappuzzo2020creating}, or learning contextualized representations with the help of contrastive learning \cite{fan2022semantics}. Recently, a method that uses column similarities on metadata and values as features for supervised classification was introduced \cite{bharadwaj2021discovering}.

\begin{figure}[t!]
    \centering
    \includegraphics[width=.6\columnwidth]{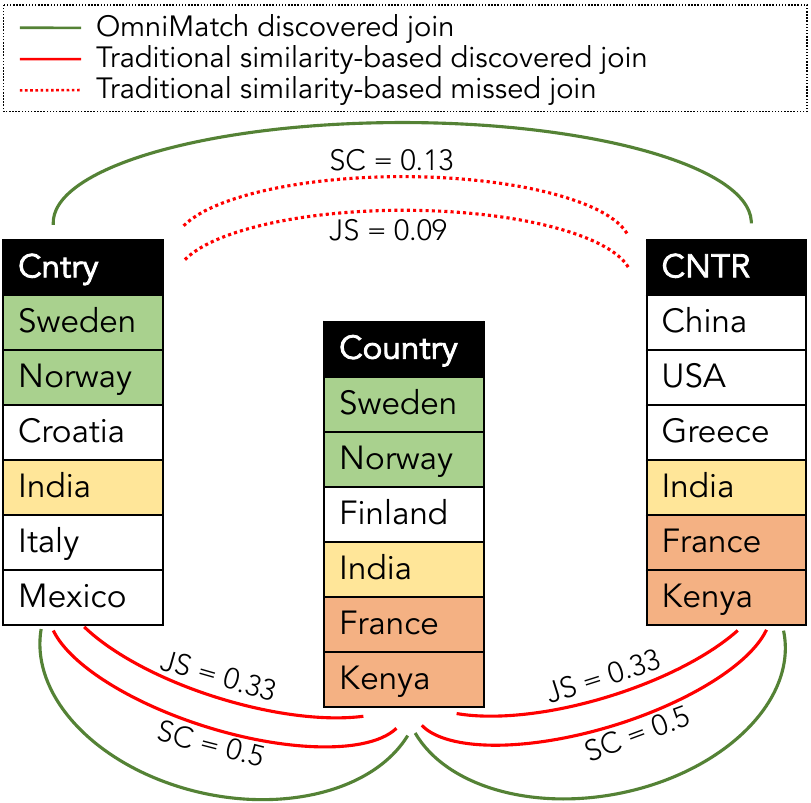}
    \caption{\method{} at work: (best viewed in color) traditional similarity-based methods vs. \method{}. If the similarity-based threshold is set to 0.3 for Jaccard Similarity (JS) or to 0.5 for Set Containment (SC), traditional methods will miss the match between columns \texttt{Cntry} and \texttt{CNTR}. Choosing these thresholds is very hard in practice as those are use-case- and dataset-dependent. \method's GNN-based method is able to discover joins using graph neighborhood information, despite the low similarity between columns, without user-provided thresholds.} 
    
    \reviewer{2: W1- First, the usage of RGCN is motivated by graph transitivity (Figure 2), but why graph transitive closure w.r.t. the filtered column pairs does not work (as it also uses filtering thresholds in Intuition of Page 2)? It is better to provide a runtime example or discussion to see the differences. $\rightarrow$ We can demonstrate the capability of RGCN through three scenarios that shed light on how it encapsulates join relations: 1) High Jaccard Similarity (JS) scores between two columns can directly suggest a join relation, 2) The differing connected columns between two columns can strengthen the evidence for join relations, and 3) Shared connected columns can stimulate the formation of join relations (Transitivity)} \reviewer{3: W1- However, it is not quite clear from the paper why RGCNs are the right choice for join graphs. Joins are not necessarily semantically transitive. A $\Join$ B and B $\Join$ C in discovery setting does not necessarily inform us about A $\Join$ C. I would suggest revisiting RGCNs and explaining why and when they can be useful for column join discovery.}
    \label{fig:graph-transitivity}
    \vspace{-4mm}
\end{figure}


However, these solutions suffer from at least one of the following issues despite their usefulness.  \emph{i}) Limited similarity metrics: these methods often choose a small and fixed set of similarity metrics to determine potential joins, limiting their flexibility in capturing diverse join scenarios.
\emph{ii}) Dependency on similarity thresholds: most existing solutions require similarity thresholds to determine potential joins based on exact value overlaps, which can lead to missed or incorrect matches when values do not perfectly overlap.
\emph{iii}) Ignoring data noise: many methods do not adequately account for noise in the data, resulting in less accurate join discovery. Data perturbations or inconsistencies are not properly handled, reducing the robustness of these methods.
\emph{iv}) Dependency on metadata: existing solutions heavily rely on clean and human-understandable metadata for join discovery. However, in practice, metadata can be noisy or even unavailable, limiting the effectiveness of these approaches \cite{madhavan2001generic,melnik2002similarity}.
\emph{v}) Need for labeled data: certain methods \cite{bharadwaj2021discovering} rely on large amounts of labeled data to train column relatedness models, which can be expensive and labor-intensive.

\para{\method: Effective Any-join Discovery} In this paper we present \method{}, a novel self-supervised approach that targets the problem of any-join discovery in tabular data repositories. \method{} effectively addresses the issues associated with existing join discovery methods in the following ways:
%
\emph{i}) \emph{Enhanced similarity metrics}: \method{} leverages a diverse suite of similarity metrics between column pairs from different datasets, enabling a more comprehensive understanding of column relatedness. 
\emph{ii}) \emph{Flexible join detection}: \method{} considers both equi and fuzzy joins by consolidating and propagating various similarity signals using a variant of \emph{Graph Neural Networks} (GNNs)~\cite{schlichtkrull2018modeling}, effectively handling diverse join conditions. 
\emph{iii}) \emph{Robustness to data noise}: by incorporating a graph-based representation that captures the inherent structure of the data, \method{} can handle noise and perturbations in the input datasets, resulting in more accurate join discovery outcomes. 
\emph{iv}) \emph{Metadata independence}: \method{} focuses more on the column content data and utilizes the column relatedness information captured in the graph, allowing it to perform join discovery even when metadata is noisy or unavailable. 
\emph{v}) \emph{Data labeling free}: \method{} employs a self-supervised learning approach by generating join examples from the original datasets, completely eliminating the need for large amounts of labeled data. This makes \method{} practical and applicable in data-scarce or labeling-challenged scenarios.

\para{Intuition} \Cref{fig:graph-transitivity} depicts three datasets with different similarity scores (Jaccard Similarity -- JS and Set Containment -- SC). The column pairs (\texttt{Country}, \texttt{CNTR}) and (\texttt{Cntry}, \texttt{CNTR}) have high similarities, while \texttt{Cntry} and \texttt{CNTR} have very low similarities. Traditional similarity-based methods rely on a user-defined threshold, often set at a low value (i.e., JS$\ge$0.09) to discover those joins, negatively affecting precision. In contrast, \method{} harnesses the power of GNNs with messaging-passing mechanisms, utilizing graph neighborhood information. This approach allows the discovery of joins that remain undetectable when using threshold-based discovery methods. By leveraging GNN, \method{} enhances the precision of join discovery without the need for a predefined threshold.

\para{Contributions} In short, the proposed \method has the following desirable properties:

\begin{itemize}[leftmargin=*]
    \item \myBoldTag{Automatic}: it takes a self-supervised approach to find equi and fuzzy joins among tabular datasets in a data repository that automatically generates positive and negative examples for self-training, using the power of GNNs. 
    \item \myBoldTag{Effective}: it decreases the number of false negatives by discovering indirect join relationships. \method{} transforms similarity signals into a graph that represents relationships among columns of different datasets. At the same time, the negative join examples used during training make \method{} robust against false positives.
    \item \myBoldTag{Extensible}: its graph modeling scheme is the first to accommodate an expandable set of well-studied similarity signals between column pairs that cover semantics via column embeddings extracted from existing deep learning approaches, value distributions, as well as set similarities.
    \reviewer{2: W3- The contribution of extensibility is limited and misleading. This work applies the RGCN network for predicting joinable tables and using more similarity metrics as input features is not a new idea. Indeed, this approach does not show how to extend an existing model with more features. If the extensibility is only to use more similarity metrics and to train a new model, it should be misleading to be claimed as "extensible", because the prediction does not have an original model to compare with, i.e., an original one to extend. This work could still claim the extensibility contribution by discussing how the model changes when more similarity dimensions are introduced. Besides extending the similarity metrics, the extensible property could be on data scales, i.e., how to generalize the trained model to new/larger datasets.}
    \item \myBoldTag {Practical}: On real-world data, \method{} achieves 14\% higher F1 and AUC scores, compared to the state-of-the-art column matching and dataset discovery methods.
\end{itemize}

%% file: 2_Problem_Definition.tex
\section{The Any-Join Discovery Problem}
\label{sec:problem}

This work addresses the problem of any-join discovery among columns from tabular datasets within a given data repository. Tabular data are abundant in every organization that maintains such repositories, which can store CSV files, spreadsheets, and database relations. Therefore, finding join relationships among their columns can better leverage the information stored in them. In the following we define the types of joins that our method focuses on.

\begin{figure*}[t!]
    \centering
    \includegraphics[width=1\textwidth]{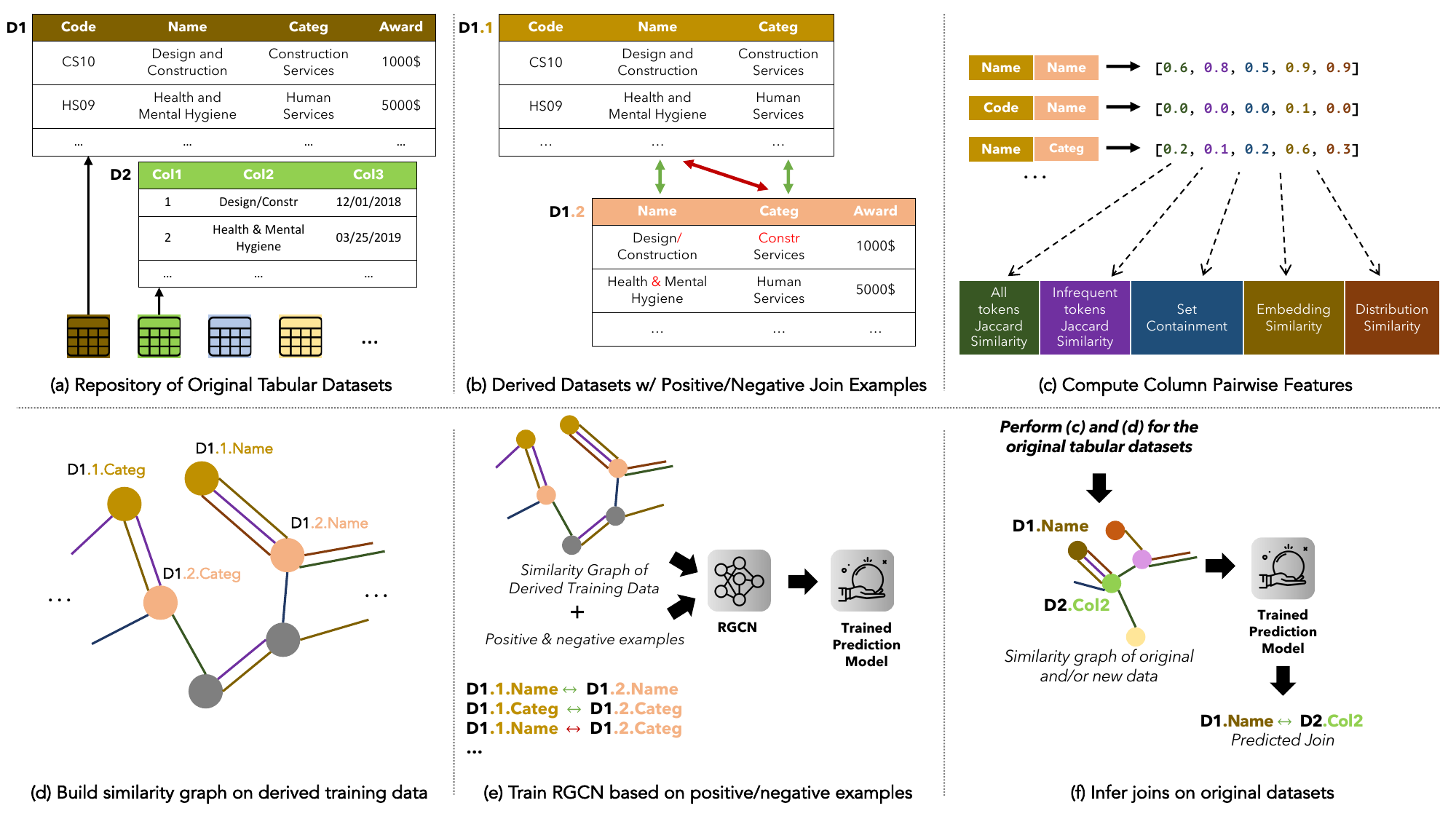}
    \caption{\method{} overview: (b) positive and negative join examples are generated in a self-supervised manner based on the original data repository shown in (a). For each positive and negative join pair, \method{} computes a set of similarity signals (c) and then constructs a similarity graph (d), which represents the most prominent column relationships among training data. The similarity graph and the join examples are the basis for producing column representations through a GNN and training a join prediction model, as shown in (e). For discovering joins, we repeat steps (c) and (d) for the original tabular datasets in the repository and use the trained model to infer joins among their columns. Best viewed in color.}
    \label{fig:overview}
\end{figure*}







\begin{definition}[\textbf{Equi-join}]
Two columns $A$ and $B$, with corresponding value sets $\mathcal{A}$ and $\mathcal{B}$, represent an \emph{equi-join} pair if \emph{i}) they share values, i.e., $\mathcal{A} \cap \mathcal{B} \not\equiv \emptyset $ and \emph{ii}) they store values from the same domain, i.e., of the same semantic type.
\end{definition}

In principle, a value overlap between two columns indicates an equi-join relationship only if their domains coincide. Pairs of columns in \autoref{fig:graph-transitivity} represent valid equi-joins since they share exact overlaps of values belonging to the same domain, i.e., country names. However, tabular datasets in a data repository come from different sources with disparate value encodings. We refer to these cases as \emph{fuzzy-joins}.

\begin{definition}[\textbf{Fuzzy-join}]
Two columns $A$ and $B$, with corresponding value sets $\mathcal{A}$ and $\mathcal{B}$, represent a \emph{fuzzy-join} pair if \emph{i}) there exists a function $h: \mathcal{A} \rightarrow \mathcal{B}$  so that they share values, i.e., $h(\mathcal{A}) \cap \mathcal{B} \not\equiv \emptyset$ and \emph{ii}) they store values from the same domain, i.e., of the same semantic data type.
\end{definition}

A fuzzy-join example is shown in \autoref{fig:similarities}, where both columns store street addresses using different formatting conventions. Fuzzy-join discovery is a challenging task since it is difficult to strictly define the function $h(\cdot)$ that transforms the values of one column to coincide with the ones of the other column syntactically; examples of such functions might drop, rearrange, or abbreviate tokens.

\para{Any-join Discovery in Tabular Data Repositories} Given a data repository consisting of a set of tabular datasets, the problem of \emph{any-join discovery} is to capture potential equi-joins and fuzzy-joins among columns belonging to different datasets stored in the repository.



%% file: 3_Approach_Overview.tex
\section{Approach Overview}
\label{sec:overview}

\autoref{fig:overview} summarizes \method's steps towards building a prediction model for any-join between columns of tabular datasets in a repository.

\noindent-- \emph{Creating training examples}: \method utilizes a dedicated \textit{dataset join-pair generator} for the datasets that reside in a given repository (\autoref{fig:overview}b) to establish the self-supervision. The created positive and negative join examples from individual tables serve as supervisions for training \method's prediction model to discover joinable relationships across tables.

\noindent-- \emph{Pairwise column feature computation}: At the core of \method we featurize all column pairs among the generated joinable datasets by computing several similarity signals that are widely used in the literature for capturing column relatedness (\autoref{fig:overview}c).

\noindent-- \emph{Column similarity graph construction}: Using the features we calculated earlier, we build a similarity graph where columns are connected with different similarity types of edges, each corresponding to a different feature (\autoref{fig:overview}d). To reduce the noise in graph construction, we propose a filtering strategy.

\noindent-- \emph{Training}: Based on the similarity graph, \method leverages the \textit{Relational Graph Convolutional Network} (RGCN) architecture, a variant of GNNs, in conjunction with the positive and negative join examples from the first step, to train a prediction model for joins (\autoref{fig:overview}e).

\noindent-- \emph{Inference on original datasets}: \method is an inductive model and can adapt to new datasets. Specifically, \method repeats the column pairwise feature computation and similarity graph construction steps for the original testing repository datasets. Applying the prediction model on this similarity graph, we can effectively infer joins among the tabular datasets residing in the repository (\autoref{fig:overview}f).

\para{Why Graph Neural Networks (GNNs)} 
The graph-based data model over the columns creates opportunities for \method to use similarity signals that go beyond the profiles of each column. Specifically, \method constructs a multi-relational~\cite{bordes2013translating} graph using columns as nodes\footnote{In this paper, we use the terms, columns and nodes, interchangeably.} and edges representing various types of ``relatedness'' between the nodes. The fact that an edge connects two columns indicates that they are similar according to a pairwise similarity metric (e.g., Jaccard Index or embedding similarity). However, using different signals to predict joinable relationships is non-trivial in such a graph. GNNs can automatically extract signals from the raw input graph through a message passing mechanism. This mechanism generates representations that aggregate diverse neighboring signals via different relations. Specifically, \method adopts the Relational Graph Convolutional Network (RGCN) model, a type of GNN that can effectively handle multi-relational data. Intuitively, the joinable relationship discovery can be seen as a learning problem over the constructed multi-relational similarity graph. The RGCN model aims to construct a new graph that consists of the same nodes (columns) but only contains edges that connect the joinable columns. This view is partially observed based on \method's self-created joinable pairs. Through its learning process, such a partial observation trains the RGCN to gradually learn how to encode signals from a column's profile and its $k$-hop neighboring columns connected via different relatedness relations (i.e., similarity metrics). Note that \method is inductive and can adapt to unseen datasets.

%% file: 4_Graph_Construction.tex
\section{Column Similarities as a Graph}
\label{sec:model}

This section discusses how \method{} builds a graph representing column relatedness to train a join prediction model. We first describe the similarity signals that \method{} considers. Then, we show how these similarities constitute the basis for building a similarity graph among columns of different tables and analyze the construction process.

\subsection{Pairwise Column Similarities} 
\label{ssec:similarities}

A main part of \method{} is figuring out how similar two columns are to find possible joins. We picked these similarity signals after many studies on column matching and related dataset discovery. Next, we explain the set of similarity signals we used in our method and why we use them.

\para{Jaccard Similarity on All Tokens}
Jaccard similarity is a widely used similarity metric to assess column relatedness. Specifically, this similarity score is calculated as the size of the intersection divided by the size of the union of the set of values included in two columns ($A$ and $B$), i.e.,  $J(A,B)={{|A\cap B|} \over {|A\cup B|}}$. Note that for computing this metric, we regard the entirety of the cell values that a column contains, i.e., we consider all tokens. Jaccard similarity is the most commonly used metric to inspect whether two columns store a considerable amount of overlapping values, which is a strong indicator of equi-join relationships \cite{deng2017silkmoth, fernandez2018aurum, bogatu2020dataset, zhang2020finding}.

\begin{figure}[t!]
    \centering
    \includegraphics[width=.6\columnwidth]{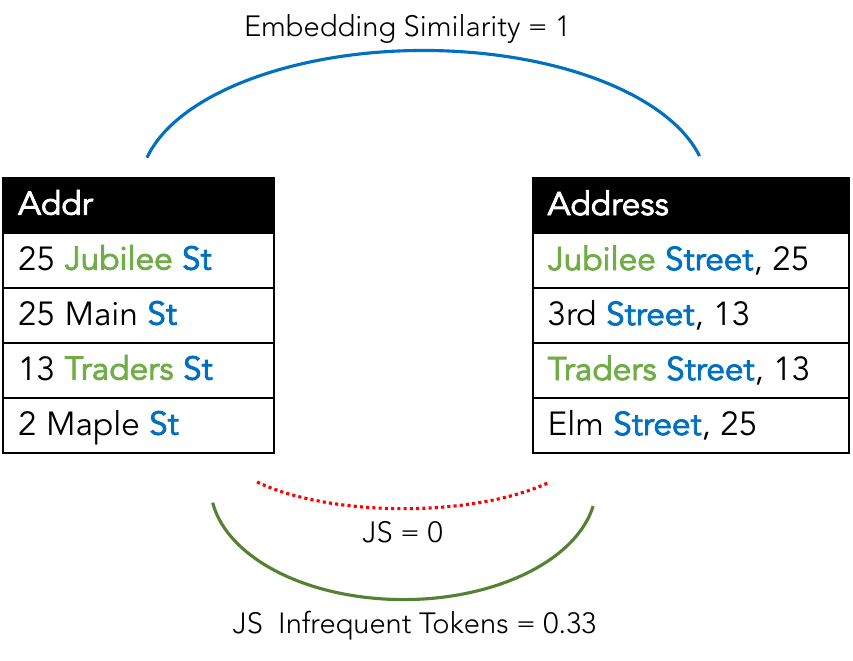}
    \vspace{-3mm}
    \caption{Using Jaccard similarity on infrequent tokens and embedding similarity on frequent tokens for capturing fuzzy-joins.}
    \label{fig:similarities}
    \vspace{-6mm}
\end{figure}

\para{Jaccard Similarity on Infrequent Tokens}
Jaccard similarity based on the complete formats of the values stored in columns strongly indicates an equi-join, yet it might be ineffective in fuzzy-joins. This is because even a slight change in the formats of values in one of the columns (e.g., \emph{St} instead of \emph{Street}) might cause the signal to be close to zero. Therefore, it is helpful to include a Jaccard similarity signal based on individual tokens stored in a column rather than on full values. To do so, \method{} includes a metric recently used in a state-of-the-art dataset top-$k$ search method \cite{bogatu2020dataset}, which we call Jaccard similarity on infrequent tokens. 

Specifically,  we first tokenize the values of each column and create a histogram of their occurrences. Then, for each value, with possibly multiple tokens, we keep as its representative the token that has the \textit{lowest frequency}. This enables us to compute Jaccard similarity on the sets of infrequent tokens stored in each column. Intuitively, a high value for this similarity signal indicates a strong relatedness between the corresponding columns since they overlap on tokens that are hardly found in their value sets. \Cref{fig:similarities} depicts an example of a fuzzy join between two columns storing addresses. Using Jaccard similarity on infrequent tokens (i.e., street names), we can capture relatedness between these two columns. On the contrary, Jaccard similarity on full values is zero. 

\para{Set Containment}
There are multiple cases where Jaccard similarity might be a weak signal of column relatedness, even if the size of overlapping values is relatively large for one of the columns. Essentially, if a column with a small value set is completely covered by another one that stores thousands of discrete values, the Jaccard similarity will be low; indeed, the size of the intersection will be relatively much smaller than the size of the union of values stored in the corresponding columns. To ameliorate this problem, several methods \cite{yakout2012infogather, deng2017silkmoth} employ \emph{set containment}. Specifically, the set containment from column $A$ to column $B$ is defined as  $|A\cap B| \over |A|$ and indicates how many unique values from $A$ are included in the intersection with $B$; a set containment of 1 indicates that those of column $B$ fully cover values of column $A$. Since this similarity measure is asymmetric, in \method{}, we choose to include the maximum set containment for a pair of columns (from one to another and vice versa). This way, we include the strongest similarity signal between the two columns. Notably, set containment is significantly effective for capturing \emph{inclusion dependencies} among columns, which is a significant step towards primary key - foreign key (PK-FK) relationship discovery \cite{zhang2010multi}.

\para{Embedding Similarity}
\method{} is designed to rely on data instances of the tables, when meta-data, such as curated column/table names or descriptions, is not available. Therefore, we compute semantic relatedness for a column of pairs by using \emph{embedding similarity} of their data instances. Value-based similarity based on pre-trained word embedding models, such as \textit{Glove} \cite{pennington2014glove} and \textit{FastText} \cite{joulin2016bag}, has been widely used in related dataset search \cite{nargesian2018table, bogatu2020dataset, dong2021efficient, khatiwada2022integrating} to capture the semantics of values stored in the columns of tabular datasets. 

In \method{}, we decided to employ value-based embedding similarity between columns by adopting the approach introduced in \cite{bogatu2020dataset}. Specifically, for each cell value in a column, we keep the token with the highest occurrence frequency based on the histogram created for computing Jaccard similarity on infrequent tokens. Next, for each such frequent token, we compute a word embedding using FastText, since it can produce representations of any given token, regardless of whether it is included in its vocabulary. Hence, this is a perfect fit for tokens containing misspellings or typos. The column representation is then computed as the mean of all embeddings of frequent tokens in the column, and the similarity between the two columns is based on the cosine similarity of their corresponding embeddings. Frequent tokens are usually representative of the column's domain. Hence, basing embedding similarity on them can strongly indicate semantic relevance between columns. For instance, in \autoref{fig:similarities} we see that embedding similarity on frequent tokens ($St$ and $Street$) suggests that both columns store values from the same domain (i.e., street addresses).

\para{Distribution Similarity}
The last signal that \method{} considers for a pair of columns is their distribution similarity. Virtually, this type of similarity is often used to capture column relatedness when their value intersection is low (i.e., Jaccard similarity is low) \cite{zhang2011automatic, nguyen2011synthesizing}, based on the observation that columns storing values from similar domains usually have relevant distributions. Distribution similarity can be beneficial when capturing synonymous terms stored in different columns, which may differ syntactically since we expect them to share similar contexts. Significantly, such a similarity signal could facilitate the discovery of fuzzy joins in \method. Consequently, in \method{}, we opted for \emph{Jensen-Shanon} (JS) divergence \cite{manning1999foundations} as the distribution similarity measure between two columns, adopting it from \cite{nguyen2011synthesizing} where it was found to be effective towards finding similar values for column matching.

Note that \method{} can be configured to compute other similarity signals due to its flexible design. Essentially, adding similarity signals in the method means adding new types of edges in the similarity graph, as discussed in the following. Therefore, \method{} can easily be modified to tailor the characteristics of the underlying datasets in a data repository by extending it to include other pairwise column similarities.

\subsection{Similarity Graph Construction}
\label{ssec:graph}

\method's pairwise column similarities can provide strong indicators of join relationships. However, relying solely on a single similarity metric can negatively affect the effectiveness of a join discovery method. As we show in \Cref{fig:graph-transitivity}, a column pair with a low JS score can still be a valid join but will be missed out if a high threshold is chosen. Moreover, some similarity measurements can become less reliable due to discrepancies in data formats, as we have discussed in \Cref{ssec:similarities}.

\para{Similarity Signals as a Graph} \method{} uses these similarity signals to construct a \emph{similarity graph}, which encodes important column relatedness information and enables \method{} to discover indirect join relationships. Specifically, columns from different datasets are transformed into nodes in a graph connected with edges of different types. Each edge type corresponds to a different similarity signal. Such a graph-based data model allows \method{} to learn $i)$ the characteristics of column profiles in join and non-join cases, \emph{ii}) whether different similarity signals contribute to a join or non-join case and \emph{iii}) whether there are graph patterns with pairwise similarity signals and column profiles that constitute a join/non-join case. 

\para{Similarity Signals \& Thresholds} Including every type of edge for each pair of nodes would result in a complete graph that is difficult to interpret and leverage towards join discovery. The most straightforward approach to filtering out edges would be to choose similarity thresholds for each similarity type. However, if we employ this graph construction technique, we might lose important column relatedness information (and graph connectivity), as it is hard to assess how suitable a value for a threshold is. For example, in \autoref{fig:graph-transitivity}, using a threshold above \texttt{0.5} would filter out all possible edges between the corresponding columns, whereas all column pairs represent a valid join relationship. 

\para{Top-\textit{k} Similarity Types per Node} To ensure high graph connectivity while accounting for different similarities, \method{} opts for a different approach: for each node in the graph, it keeps only the top-\emph{k} edges per node and per type based on the value of the corresponding similarity signal. Essentially, for each node (i.e., column), \method{} keeps the edges that represent the most prominent join relationships with other nodes. For instance, if we set $k = 1$ in \autoref{fig:graph-transitivity}, then the only edges that will be kept are the ones between the \emph{Cntry-Country} and \emph{Country-CNTR} pairs. Most importantly, in \method{}, the value of $k$ is automatically selected based on the validation set during the training of the join prediction model, as discussed in  \Cref{sec:exp}. As a result, the edges of the similarity graph that \method{} constructs using the aforementioned top-$k$ edges represent candidates of potential join relationships between the corresponding columns. 
\reviewer{1: D4-to explain why this is the case and why these are good for the propagation that will happen through the graph modeling. $\rightarrow$ We can showcase that RGCN is a generalized function of $f_\theta(x_i, x_j, e_{i,j})$}
Yet, the graph is not guaranteed to contain edges connecting every possible true column join pair (e.g., \emph{Cntry} and \emph{CNTR} share no edges for $k = 1$). As we see in the following section, \method{} tackles this issue by taking advantage of transitive paths in the similarity graph, to capture joins indirectly.

\reviewer{3: W2: [Novelty] The paper emphasizes the drawbacks of using threshold for search. The existing top-K search techniques [a,b] are indeed alternatives to threshold-based search and does not suffer when the scores are low in the repository.
[a] A Sketch-based Index for Correlated Dataset Search, Santos, ICDE, 2022.

W3: [Related Work] The paper talkes about fuzzy- and equi-joins. Existing research on semantic join discovery [a] models the join discovery problem as a search problem on bipartite graph matchings of a query column and candidate columns. The choice of similarity function is flexible enough to handle both semantic and fuzzy joins.
[b] KOIOS: Top-k Semantic Overlap Set Search, Mundra et al., ICDE, 2023.

W4: [Novelty] The join graph construction follows the existing works [c] (with new more recent measures added as edges) and can be summarized in the write-up.
[c] Linking Datasets Using Word Embeddings for Data Discovery, Fernandez, ICDE, 2018. }

%% file: 5_Model_Training.tex
\section{Graph Model Training}
\label{sec:training}

We denote the constructed similarity graph as $\mathcal{G}=(\mathcal{V}, \mathcal{E}, \mathcal{R})$ with nodes (columns) $v_i\in \mathcal{V}$ and edges $(v_i, r, v_j)\in \mathcal{E}$, where $r\in \mathcal{R}$ is a relation type indicating one of five similarity relation types defined in Section~\ref{ssec:similarities}. 
In this section, we discuss how \method{} leverages the graph $\mathcal{G}$ to learn column representations with GNNs. We begin by exploring the process of creating the initial column features. Subsequently, we employ the message passing paradigm of GNNs to calculate the aggregated column representations and provide a detdailed illustration of using RGCN in \method. Then, we explain how \method{} automatically creates positive and negative column joins for training and how to use different loss functions to guide training. Finally, we discuss how to do inference.

\subsection{Initial Column Features}
\label{ssct:column_feature}
We describe a column with a collection of identified features that better represent its characteristics \cite{abedjan2015profiling}. We denote the initial feature vector for a column $i$ as $\mathbf{x}_i \in \mathbb{R}^{d_f}$, where $d_f$ denotes the feature dimension.  
%
%
Specifically, for each column, we use a simple profiler that summarizes statistical information about the values of a given column. We do so since more complex information about the column contents is captured by different types of edges among the nodes in the similarity graph. To this end, we make use of the column profiling component from Sherlock \cite{hulsebos2019sherlock} by computing statistics falling into the following two categories:

\parait{-- Global statistics} Those include aggregates on high-level characteristics of a column, e.g., the number of numerical values. We use the implementation\footnote{\url{https://github.com/mitmedialab/sherlock-project/blob/master/sherlock/features/bag_of_words.py}} from Sherlock \cite{hulsebos2019sherlock}.

\parait{-- Character-level distributions} For each of the 96 ASCII characters that might be present in the corresponding values of the column, we save character-level distributions. Specifically, the profiler counts the number of each such ASCII character in a column and then feeds it to aggregate functions, such as \emph{mean}, \emph{median} etc. Our implementation is based on the original character-level distributions features\footnote{\url{https://github.com/mitmedialab/sherlock-project/blob/master/sherlock/features/bag_of_characters.py}} in Sherlock \cite{hulsebos2019sherlock}.


\subsection{Column Representation Learning via Message Passing}
\label{ssec:message-passing}
Next, we build upon the message-passing architecture of GNN, specifically RGCNs, to capture necessary similarity signals within the graph and refine the representation of columns. 

\subsubsection{The message passing paradigm for GNNs}
The message passing paradigm follows an iterative scheme of updating node representations based on the aggregation from neighboring nodes. 
Suppose $\mathbf{h}^{(\ell)}_{i}$ represents the node representation for column $i$ at iteration $\ell$, then the paradigm composes four parts:
\begin{enumerate}
    \item Initialization: $\mathbf{h}_i^{(0)}= f_{\theta_1}(\mathbf{x}_i), \forall v_i \in \mathcal{V}$. For each node $i$, we initialize its node representation $\mathbf{h}_i^{(0)}$ as a function of the feature vector defined in Section~\ref{ssct:column_feature}.
    \item Message computation: $\mathbf{m}_{i\leftarrow j}^{(\ell)} = \phi_{\theta_2^{(\ell)}}( \mathbf{h}_j^{(\ell-1)},  \mathbf{h}_i^{(\ell-1)}, \mathbf{e}_{i,j}^{(\ell-1)})$. Function $\phi_{\theta_2^{(\ell)}}(\cdot)$ parameterized by $\theta_2^{(\ell)}$ computes a message from each neighboring node $j$ to the central node $i$. Here, $\mathbf{e}_{i,j}$ denotes edge information between nodes $i$ and $j$, which contains the information of a specific relation type.
    \item Neighbor aggregation: $\mathbf{m}_i^{(\ell)} = \psi_{\theta_3^{(\ell)}}( \{\mathbf{m}_{i\leftarrow j}^{(\ell)} | j\in \mathcal{N}_i\})$. This step aggregates the messages received from all the neighboring nodes defined by $\mathcal{N}_i$ to form a comprehensive message for node $i$. $\psi_{\theta_3^{(\ell)}}(\cdot)$ is the function parameterized by $\theta_3^{(\ell)}$ that aggregates messages.
    \item Message transformation: $\mathbf{h}_i^{(\ell)} = f_{\theta_4^{(\ell)}}(\mathbf{h}_i^{(\ell-1)}, \mathbf{m}_i^{(\ell)})$. Function $f_{\theta_4^{(\ell)}}(\cdot)$ parameterized by $\theta_4^{(\ell)}$ transforms the aggregated information into an updated representation for node $i$.
\end{enumerate}

In summary, the GNN message-passing paradigm initializes node representations, computes messages between neighboring nodes, aggregates these messages, and transforms the aggregated information to update node representations in an iterative manner.  a column gains the ability to receive a greater number of relevant messages from its neighbors at the $L$th-hop. This enables us to delve into high-order connectivity information and enhance our understanding of intricate relationships within the data. Such high-order connectivities are crucial to encode the similarity signal to estimate the joinable score between two columns. 
The parameters $\theta_1$, $\theta_2^{(\ell)}$, $\theta_3^{(\ell)}$, and $\theta_4^{(\ell)}$ are adjustable based on different GNN architectures and can be learned during the training of GNN.

\subsubsection{Relational Graph Convolutional Network (RGCN)}
In \method, we leverage the power of the RGCN model to effectively capture multi-relational and multi-hop neighboring features.

\para{Node feature initialization}
We set the initial value of $\mathbf{h}_i^{(0)}$ as $\mathbf{x}_i$ in $\mathbb{R}^{d_f}$, with an empty parameter set $\theta_1$.

\para{Message Computation}
Intuitively, the neighboring columns in a similar graph can give more clues about the semantic meaning of a column. We build upon this basis to encourage message feature propagation between linked columns under different types of similarity relations as follows. In \method{}, we use linear transformations as the encoding function:
\begin{equation}
    {\mathbf{m}_{i\leftarrow j}^r}^{(\ell)} = \frac{1}{|\mathcal{N}_i^r|} (\mathbf{W}_r^{(\ell)} \mathbf{h}_j^{(\ell-1)} + \mathbf{b}_r^{(\ell)}) + \frac{1}{\sum_{r \in \mathcal{R}} |\mathcal{N}_i^r|} (\mathbf{W}_0^{(\ell)} \mathbf{h}_i^{(\ell-1)} + \mathbf{b}_0^{(\ell)}),
    \label{eq:message}
\end{equation}
where $\mathbf{W}_r^{(\ell)} \in \mathbb{R}^{{d_{h}^{(\ell)}} \times d_{h}^{(\ell-1)}}$ is a weight matrix for relation $r$, which transforms a column feature vector of dimension $d_{h}^{(\ell-1)}$ to a hidden dimension $d_{h}^{(\ell)}$. There is also a different weight matrix $\mathbf{W}_0 \in \mathbb{R}^{{d_{h}^{(\ell)}} \times d_{h}^{(\ell-1)}}$ that helps preserve some of the original information (residual connection). So, we have $\theta_2^{(\ell)} = \{\mathbf{W}_r^{(\ell)}, \mathbf{b}_r^{(\ell)}, \mathbf{W}_0^{(\ell)}, \mathbf{b}_0^{(\ell)} \}$. $\mathbf{b}_r^{(\ell)} $ and $\mathbf{b}_0^{(\ell)} $ are the bias vectors. 
$\mathcal{N}_i^r$ stands for the set of neighboring columns of $i$ under relation $r \in \mathcal{R}$ and $\sum_{r \in \mathcal{R}} |\mathcal{N}_i^r|$ indicates the total number of neighbors under all types of similarities. Thus, the coefficient scalar controls the number of messages being propagated based on the degrees of the node under each relation.

\para{Neighbor Aggregation}
In the aggregation stage, messages from neighboring columns are passed to the target column via different types. This helps refine the understanding of our target column $i$:
\begin{equation}
    \mathbf{m}_i^{(\ell)} = \sigma(\sum_{r \in \mathcal{R}}\sum_{j\in \mathcal{N}_i^r} {\mathbf{m}_{i \leftarrow j}^r}^{(\ell)} ),
    \label{eq:agg}
\end{equation}
After computing the aggregated specific messages, we sum the messages from all types and pass the output to a non-linear function $\sigma(\cdot)$. Here $\mathbf{m}_i^{(\ell)}$ denotes the representation of column $i$ after aggregating $\ell$ column propagation layers. We use sigmoid as the activation function $\sigma(\cdot)$, since it allows messages to encode positive signals and filter the negative ones. 

\para{Message Transformation} We use the residual connection without any additional parameters to update the node representaion. In addition to the messages propagated from the neighbors under different similarity channels, we consider the self-connection of $i$, which retains the information of the original column features:
\begin{equation}
    \mathbf{h}_i^{(\ell)} = \mathbf{h}_i^{(\ell-1)} + \mathbf{m}_i^{(\ell)}.
    \label{eq:update}
\end{equation}
At the $L$-th layer, the node representations are $\mathbf{h}_i^{(L)}, \forall v_i \in \mathcal{V}$.

\subsection{Generating Training Examples}
\label{ssec:generating}
\reviewer{1: D1-It is not clear how are join examples generated from the original datasets and why the proposed process is justified and can help with training. D5-The section about training examples generation (5.2) needs to be totally rewritten and more importantly justified on why this a good idea. It is not very clear how the training examples are obtained and why it is correct that these training data is good for training.
}
Training our prediction model requires join (positive) and non-join (negative) labels. To do so, \method{} takes a self-supervised approach, leveraging positive and negative join examples that are automatically generated from the tabular data in the repository. Specifically, for each table in the input, \method{} adopts a join pair fabrication process \jiani{we should expand the pair fabrication process.}, similar to the ones described in \cite{nargesian2018table, zhu2019josie, koutras2021valentine}. Specifically:

\begin{itemize}[leftmargin=*]
    \item We randomly pick some columns from the input table that the derived pair of datasets will share.
    \item Then, we split the original dataset's rows into two randomly overlapping sets. Consequently, we create a pair of datasets with a random number of columns and rows.
    \item To simulate fuzzy-joins, we randomly perturb the data values of one of the two created datasets. We do so only for instances belonging to columns that are shared among the generated tables. To perturb the data values, we either \emph{i}) insert random typos based on keyboard proximity (e.g., \emph{science} becomes \emph{scienxe}) or \emph{ii}) use common alternative values formats for specific column cases (e.g., dates, money amounts, street addresses, etc.).
\end{itemize}

\reviewer{3: W6: [Learning Algorithm] Why generating training data requires selecting subsets of columns and rows and perturbing data and adding more noise instead of using original columns and their links which seem to provide an abundant training data? }

\noindent Based on the above join generation process, we create a pair of joinable tables for each original dataset in the repository (\autoref{fig:overview}b). The columns that join in these pairs are used as positive training examples, while the rest of the column combinations between the two tables are regarded as negative join examples. Note that with this generation process, the derived pairs will share joins of various overlaps and fuzziness, ensuring that our model is effective is several join scenarios.  
\reviewer{3: W5: [Learning Algorithm] The paper does not explain how negative examples are generated. Is any pairs of columns that is not connected in the graph considered as a negative example? If so, this would probably mean there exists an imbalance of negative and positive examples. Moreover, it is common in representation learning to focus on hard negative examples rather than random ones. }

\subsection{Loss Functions}
\label{ssec:loss}
To refine the column representations produced from the RGCN, \method{} leverages the automatically created positive and negative join examples to train a prediction model. \reviewer{1: D6-This is confusing. Are the authors trying to build a prediction model to decide if two column can be joined or not, or are they after a better column representations that will be then used by a prediction model? This should be rewritten in a more systematic and clear way.} In what follows, we describe two alternative training procedures that are characterized by different loss functions.

\para{Training with cross-entropy loss} In this training procedure, the model's goal is to optimize the following \emph{cross-entropy} loss function:

\begin{align}
\begin{split}
    \mathcal{L} &= - \sum_{(A,B) \in \mathcal{J}}{w_p\cdot\log{\sigma(sim(\mathbf{h}_A^{(L)}, \mathbf{h}_B^{(L)}))}} \\ 
    &\quad - \sum_{(A,B) \in \mathcal{NJ}}{\log(1-\sigma(sim(\mathbf{h}_A^{(L)}, \mathbf{h}_B^{(L)})))},
\end{split}
\end{align}

\noindent where $\sigma(\cdot)$ is the sigmoid function, while $\mathcal{J}$ and $\mathcal{NJ}$ are the sets of positive and negative column join examples. Notably, the parameter $w_p$ is the weight we use to balance the positive and the negative examples, which we set as the ratio of negative to positive join examples in training. The similarity scores are computed by feeding pairs of RGCN-produced column representations to a \emph{Multi-layer Perceptron} (MLP), whose parameters are also learned during training to give correct predictions. With this model training, we aim to compute column representations (using RGCN), so we can build a similarity function (through MLP) that scores join examples higher than non-join ones.

\para{Training with triplet margin loss} An alternative for proceeding with training is using the triplet margin loss function:
\begin{align}
\begin{split}
\mathcal{L} = \sum_{(A, B^+, C^-)}\max\{ & d(\mathbf{h}_A^{(L)}, \mathbf{h}_{B^+}^{(L)}) - d(\mathbf{h}_A^{(L)}, \mathbf{h}_{C^-}^{(L)})
\\ & + \text{margin}, ~ 0\},
\end{split}
\end{align}
\noindent where $d(\cdot, \cdot)$ is a vector distance function, and $margin$ is a positive value. For each column, we consider one column that joins (denoted by $+$) and all others that do not join (denoted by $-$) based on the generated dataset pairs. Intuitively, training to minimize the triplet margin loss helps the RGCN learn to bring the representations of columns that join, closer than the ones that do not. 
\jiani{Christos, please double check the meaning of the above paragraph -- how to define the triplet margin loss.}

\reviewer{2: W1- the loss function aims to distinguish joinable and non-joinable column pairs from similarity measures. However, transitivity aims to classify the non-joinable column pairs (by measures) to joinable pairs because they share the same domain (Def 2 in Section 2.2). It seems that they have conflict targets. Also, Section 5.3 shows two different alternative loss functions, but how they are used in practice is not clear (since we may not train a model by two different target at the same time).}




\begin{figure}[t!]
    \centering
    \includegraphics[width=\columnwidth]{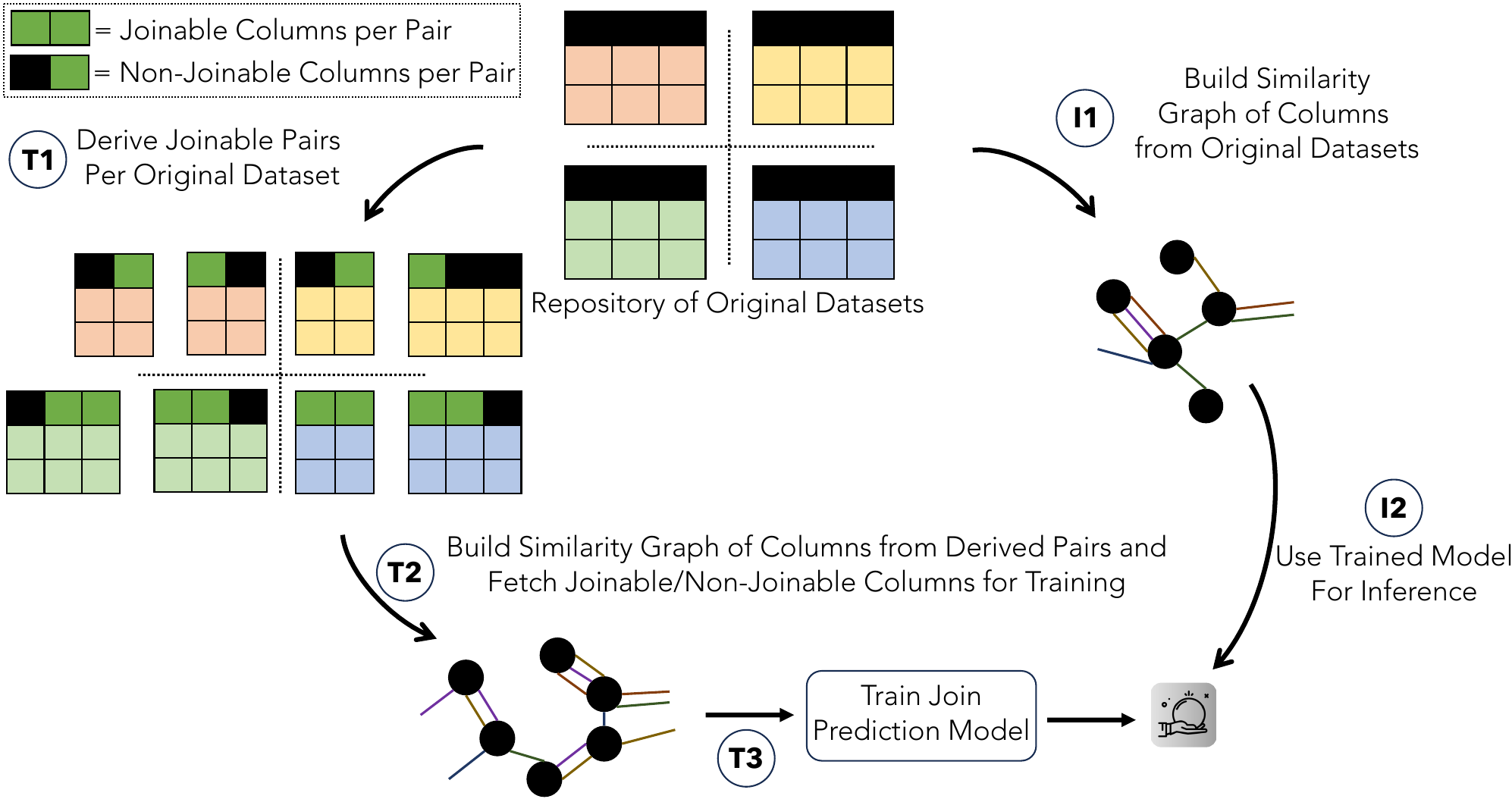}
    \caption{For training, \method{} fabricates pairs of joinable datasets (T1) from each original one in the repository to build a similarity graph (T2) for training the join prediction model (T3). For inference, \method{} constructs the similarity graph of the columns stemming from the original datasets (I1) and uses the trained model for inference on it (I2).}
    \label{fig:training-inference}
\end{figure}

\subsection{Training and Inference of Join Predictions}

Figure \ref{fig:training-inference} summarizes \method's training and inference procedures. We derive a pair of joinable pairs for each original dataset in the repository, as discussed in Section \ref{ssec:generating}. Based on the derived tables, our method first computes all pairwise column similarities and constructs the similarity graph. Then, the join prediction model training process is applied to the constructed graph, where learning is guided by one of the two loss functions, as described in Section \ref{ssec:loss}. 

While the join-prediction model-training process occurs on the derived dataset pairs, our objective is to discover joins among the columns of the original datasets in the data repository. To this end, \method{} builds the similarity graph based on the pairwise similarities of columns belonging to the original tabular datasets (right part in Figure \ref{fig:training-inference}). Based on the connectivity information of this graph, the trained RGCN model can be straightforwardly applied to retrieve the representations of columns: message aggregation takes place once to infer the column embeddings based on the weight matrices learned during training. As a last step, \method{} uses the column representations to produce a joinability score between each pair of columns coming from different datasets in the repository; the joinability score depends on the loss function used to guide the learning process (Section \ref{ssec:loss}).

\reviewer{1: D8- After going through Section 5, it is clear that it needs to be totally overhauled. The hardest one to follow is the last paragraph explaining the inference. It is way too dense and requires a lot of efforts to be understood.}
\reviewer{2: W1- Third, in Inference of Section 5.4, it is not intuitive to apply trained model to new datasets. It may be a narrow application to new datasets due to the different graph topologies for training RGCN over any two pairs of datasets, although similar domains may exist in the same data repo.}

%% file: 6_Experimental_Evaluation.tex
\section{Experimental Evaluation}
\label{sec:exp}

In this section, we present a comprehensive set of experiments that showcase the effectiveness of \method. First, we describe the join discovery benchmarks and baseline methods against which we evaluate our method. Then we provide the experimental results that demonstrate \emph{i}) the gains in effectiveness with respect to state-of-the-art methods when using \method, \emph{ii}) how \method's prediction model compares to using other models and \emph{iii}) how different similarity signals are related to the model's effectiveness. In addition, we provide execution times for the different steps of our method. We summarize our main results as follows.

\begin{itemize}[leftmargin=*]
    \item \method{} is considerably more effective than state-of-the-art column matching and column representation methods.
    \item We showcase that utilizing only one similarity signal reduces \method's effectiveness. The degree of reduction depends on the characteristics of the underlying datasets.
    \item \method's choice of using RGCNs for leveraging the set of similarity signals is superior to using alternative ML models.
\end{itemize}

\subsection{Experiment Setup}
\label{ssec:setup}

\para{Datasets}
We construct two realistic join benchmarks to properly evaluate the effectiveness of \method{} and the other methods. \autoref{tab:benchmarks} summarizes the statistics of both benchmarks. We explored the New York City OpenData\footnote{\url{https://opendata.cityofnewyork.us/}}, specifically the \emph{City Government} and \emph{Culture Recreation} tabular data repositories. The City Government benchmark consists of 110 tables derived by 11 denormalized tables using techniques of  \cite{nargesian2018table, koutras2021valentine}, i.e., horizontal and vertical partitions. Similarly, the Culture Recreation benchmark consists of 120 tables derived from 12 denormalized tables. Most columns in both benchmarks store mainly categorical and text data. At the same time, a few cases of numerical data are mostly distinguishable based on their value sets, i.e., with minimal/empty overlaps.

\para{Ground Truth} To measure effectiveness, we manually annotated column join relationships (both equi and fuzzy ones) among the corresponding base tables of the benchmarks. Based on these annotations, we automatically generated the ground truth for column pairs among all tables included in both benchmarks. Nonetheless, to secure the validity of the captured fuzzy join relationships in the ground truth, we manually inspected their correctness to avoid false positives. 

\para{State-of-the-art Baselines}
We compare \method{} against the two best-performing column matching methods, according to a recent study \cite{koutras2021valentine}, and the state-of-the-art contextualized column representation method for capturing relatedness among columns, described below.

\noindent-- \emph{COMA} \cite{do2002coma} is a seminal matching method that takes into consideration multiple similarity scores, from both metadata and data instances \cite{engmann2007instance}. COMA's effectiveness relies on processing these similarity signals from simple metrics to decide on possible column matches. In our evaluation, we make use of the COMA 3.0 Community Edition.
\begin{table}[t!]
\centering
\begin{tabular}{l|c||c||c||c}
\toprule\bottomrule
\textbf{Benchmark} & \textbf{\#Tab.} & \textbf{\#Col.} & \begin{tabular}[c]{@{}c@{}}\textbf{\#Equi-}\\ \textbf{Joins}\end{tabular}  & \begin{tabular}[c]{@{}c@{}}\textbf{\#Fuzzy-}\\ \textbf{Joins}\end{tabular} \\ \toprule\bottomrule
\textbf{City Government} & 110 &  703 & 1451 & 128 \\ \hline
\textbf{Culture Recreation} & 120 & 687 &  1254 & 256 \\
\toprule\bottomrule
\end{tabular}
\vspace{1mm}
\caption{Statistics of the evaluation benchmarks. `Tab.' stands for `Table' and `Col.' stands for `Column'. }
\label{tab:benchmarks}
\vspace{-6mm}
\end{table}

\noindent-- \emph{Distribution-Based (DB) Matching} \cite{zhang2011automatic} is an instance-based column matching method. The method constructs clusters using the \emph{Earth Mover's Distance} (EMD) to capture relatedness among columns of different tabular datasets. During cluster refinement, the method considers exact value overlaps between column pairs to avoid false positives. To include the DB matching method in our experiments, we use the implementation provided by Valentine \cite{koutras2021valentine}.

\noindent-- \emph{Starmie} \cite{fan2022semantics} is a state-of-the-art dataset discovery method. It contextualizes column representations (embeddings) to facilitate \emph{unionable table search} in \emph{data lakes}. The method employs a multi-column table encoder that serializes instances from tables to feed them into a pre-trained \emph{Language Model} (LM) (specifically, the authors use RoBERTa \cite{liu2019roberta}). Starmie uses contrastive learning \cite{chen2020simple} to produce column representations that capture dataset relatedness. In our evaluation, we use Starmie, as shared in a public repository \footnote{\url{https://github.com/megagonlabs/starmie}}, to produce contextualized column representations for the datasets in the input. We then compute the pairwise cosine similarity of the column embeddings among different datasets. To produce the best results for Starmie, we fine-tuned its parameters for both join benchmarks. 

\noindent-- \emph{DeepJoin} \cite{deepjoin} proposes a state-of-the-art deep learning model for dataset discovery, which leverages a pre-trained LMs (the authors use MPNET \cite{song2020mpnet} since it produces the best results), similarly to Starmie \cite{fan2022semantics}, in order to produce fine-tuned column representations for \emph{joinable table search} in \emph{data lakes}. Specifically, DeepJoin serializes columns as sentences by concatenating their values. These sentences are then fed to a \emph{sentence transformer} \cite{reimers-2019-sentence-bert} model to produce initial vector representations of the corresponding columns. To fine-tune them for joinable dataset search, DeepJoin trains an embedding model based on a set of positive join pairs, and towards minimizing the \emph{multiple negative ranking loss}; the set of positive join pairs is computed based on a similarity join method of choice and a high threshold to ensure lower numbers of false positives. For the needs of our evaluation, we train the DeepJoin model based on the Sentence-BERT\footnote{\url{https://www.sbert.net/}} library, to produce column  representations. As in the case of Starmie, we use pairwise cosine similarity as the joinability score between two columns, while we fine-tune DeepJoin's parameters to get the best results; positive training pairs are generated based on pairwise cosine similarity of initial column representations ($\geq$ 0.9 to ensure high true positive rates).

\begin{figure*}[t!]
    \centering
    \hspace{3mm}\begin{minipage}{\columnwidth}
            \includegraphics[width=.8\columnwidth]{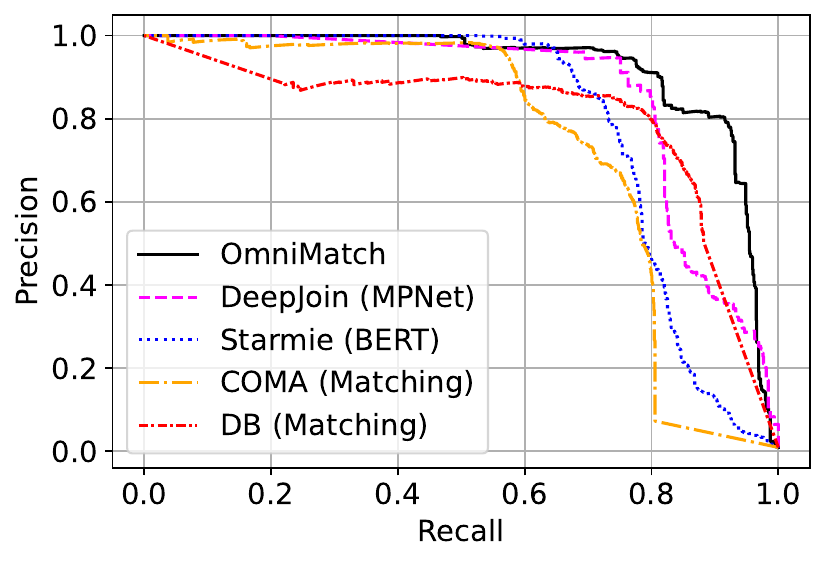}
            \vspace{-2mm}
            \subcaption{City Government}
            \label{fig:comparison_cg}
    \end{minipage}
    \begin{minipage}{\columnwidth}
             \includegraphics[width=.8\columnwidth]{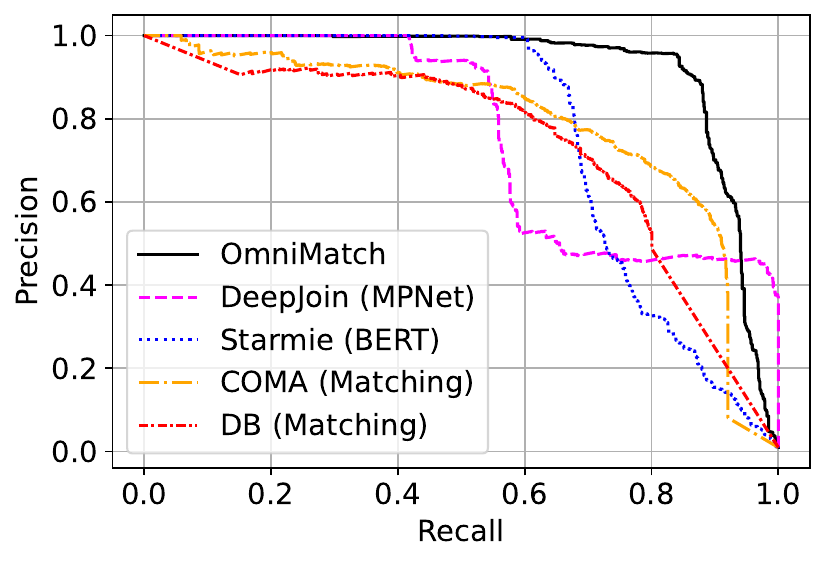}
             \vspace{-2mm}
             \subcaption{Culture Recreation}
             \label{fig:comparison_cr}
    \end{minipage}
    \vspace{-2mm}
    \caption{Effectiveness comparison of \method{} with the state-of-the-art methods for various similarity thresholds.}
    \label{fig:comparison}
    \vspace{-2mm}
\end{figure*}

\para{Other ML Predictive Methods}
We also evaluate the strength of \method's graph model and RGCN architecture by comparing it to other straightforward column join prediction models that make use of the same features (pairwise similarities) but do not take the graph information into account. Namely:

\noindent-- \emph{Random Forest} considers only the column pairwise similarities and the positive and negative join training examples that our method computes to train a binary classification method, similar to \cite{bharadwaj2021discovering}. For our experiments, we use the random forest implementation from \texttt{sklearn} \cite{pedregosa2011scikit} Python toolkit, for both training and inference, with \texttt{100} decision tree classifiers.

\noindent-- \emph{MLP} uses the same information as the Random Forest baseline but feeds them to a shallow Multi-Layer Perceptron (MLP) binary classification model. Specifically, we use an MLP with one hidden layer, which takes input as the pairwise similarities and learns to predict joins between the corresponding columns.

\para{Measuring Effectiveness}
We use \emph{Precision-Recall (PR) curves} to evaluate the effectiveness of \method{} and the other baseline methods based on the final join prediction scores for each column pair among different datasets in the benchmarks. PR curves are suitable for illustrating effectiveness results when there is an imbalanced distribution of labels in the test set. Indeed, in our case, the number of non-joinable column pairs is significantly higher than the number of joinable ones for both benchmarks, as happens in every real-world data repository. A significant advantage of using PR curves is that we can observe effectiveness for varying similarity thresholds, thus making the presentation non-biased. PR curves can help us observe how different similarity thresholds affect a method's performance; stable precision for increasing recall values means that the method's effectiveness is robust to different similarity thresholds. We also report the best F1 and PR-AUC scores to summarize the results shown in PR curves.

\para{Tuning \method} We configure \method{} by running experiments when varying the model's parameters. By doing so, we came to the following conclusions.

\noindent-- \emph{Graph Construction:} We trained \method's join prediction model for different values of top-$k$ edges that we consider in the graph for each node and similarity signal to assess changes in effectiveness. Our results showed that using values greater than \texttt{5} did not improve our model's effectiveness. To automatically decide on the value of $k$ that gives the best results for each benchmark, we use a validation set that we exclude from the training column pair samples.

\noindent-- \emph{Number of RGCN Layers:} We evaluated how the number of layers (i.e., range [\texttt{1, 3}]) used for training the RGCN affects \method's performance. Our results showed that using two layers provides the highest effectiveness gains, meaning that \method's model benefits from looking one hop away from each node (column). This verifies our intuition that leveraging transitivity in the similarity graph improves the quality of the join predictions.

\noindent-- \emph{Number of Epochs:} We trained \method{} for several epochs and used loss curves with a \texttt{90:10} training/validation data split. Notably, using more than \texttt{30} epochs does not incur considerable changes in the training/validation losses. Therefore, for the rest of the experiments, we train \method{} for \texttt{30} epochs; the same stands for the Random Forest and MLP baselines.

\noindent-- \emph{Dimension of Embeddings:} We assessed the influence on \method's effectiveness when producing column representations of varying dimensionality through the RGCN model. We ran experiments with \{\texttt{32, 64, 128, 256, 512}\} dimensions and found that column embeddings of 256 dimensions produce the best results.

\noindent-- \emph{Initial Node Features:} We evaluated how the initial node features we use for training the RGCN affect the performance of \method. Instead of using the proposed node features, we generated random feature vectors for each node of the same length as the RGCN's dimension of embeddings. Results verified the effectiveness of our node feature initialization process, as we observed a decrease of more than $10\%$ in terms of PR-AUC scores when using randomized initial node features. 

\noindent-- \emph{Loss Function:} As we discussed in \Cref{sec:training}, our training process can be guided using two different loss functions: \emph{i}) cross-entropy loss and \emph{ii}) triplet margin loss. Thus, we evaluated the effectiveness of the prediction model when employing a different loss function in both benchmarks. Notably, the results show that using triplet margin loss can greatly improve the effectiveness as opposed to the cross entropy loss; its ability to bring closer column representations of joinable pairs while setting apart the ones of non-joinable pairs helps \method{} to better distinguish between the two cases.

\para{Implementation details} 
For training, we use the Adam optimizer \cite{kingma2014adam} with a learning rate of \texttt{0.001}, while we use an MLP of one hidden layer when employing \method{} with a cross-entropy loss. \method{} is implemented in Python; for implementing the RGCN model we used the Deep Graph Library (DGL) \cite{wang2019deep} on top of PyTorch. We use an AMD EPYC 7H12 Linux machine with 128 2.60GHz cores and an NVIDIA A40 GPU.

\subsection{Comparison to State-of-the-Art Baselines}

In \autoref{fig:comparison}, we show how \method{} compares against the state-of-the-art methods (\Cref{ssec:setup}) in terms of effectiveness using Precision-Recall curves. First, our method significantly outperforms the baselines since it can consistently provide high precision values even for recall values close to \texttt{0.8}. Essentially, our method achieves high precision no matter the similarity threshold (except for very low ones), thus securing the quality of the returned joins. Interestingly, the column matching methods (COMA and DB) give low precision even for recall values that are not high, i.e., when the similarity thresholds are high. 
The reason is that these methods rely on a limited set of similarity signals based on data instances, which do not account for value semantics and syntactic differences, leading to false join predictions. Their results get worse in the Culture Recreation benchmark due to its more difficult join cases among column pairs of different datasets.
\begin{table}[!t]
\centering
\resizebox{\columnwidth}{!}{
\begin{tabular}{l|c||c|c||c||c}
\toprule
\multicolumn{6}{c}{\textbf{Best F1 Scores}}\\
\bottomrule
\textbf{Benchmark} & \textbf{\method} & \textbf{Starmie} & \textbf{DeepJoin}& \textbf{COMA} & \textbf{DB}  \\ \toprule
\bottomrule
\textbf{City Government} & \textbf{0.857} & 0.781 & 0.819 & 0.720  & 0.803\\ \hline
\textbf{Culture Recreation} & \textbf{0.894} & 0.759 & 0.681 &  0.744 & 0.708 \\
\toprule
\multicolumn{6}{c}{\textbf{PR-AUC Scores}}\\
\bottomrule
\textbf{Benchmark} & \textbf{\method} & \textbf{Starmie} & \textbf{DeepJoin} & \textbf{COMA} & \textbf{DB}  \\ \toprule
\bottomrule
\textbf{City Government} & \textbf{0.920} & 0.798 & 0.820 & 0.733  & 0.760\\ \hline
\textbf{Culture Recreation} & \textbf{0.921} & 0.765 & 0.763 & 0.786 & 0.680 \\
\bottomrule
\end{tabular}
}
\caption{Best F1 and PR-AUC scores comparison of \method{} and the state-of-the-art baselines.}
\vspace{-20pt}
\label{tab:f1_auc_sota}
\end{table}

On the other hand, Starmie, with its contextualized column representations, does not deliver high precision for recall values above \texttt{0.6}. This mainly happens due to the counter-intuition behind contextualized column representations and join discovery: columns that join among different columns do not necessarily share similar contexts. In addition, the training examples produced by Starmie do not account for value discrepancies (i.e., fuzzy joins). Similarly, DeepJoin embeddings entail low precision for high recall, especially in the case of the Culture Recreation benchmark. This is mainly due to the positive and negative pairs on which the model is trained, which are not guaranteed to be accurate. On the contrary, \method{} avoids this issue by relying on a training example generation that ensures true positive and negative pairs (Section \ref{ssec:generating}). Furthermore, challenging join cases, where value overlaps are relatively small, are difficult to be captured by DeepJoin, since it cannot propagate various similarity signals as our graph model.

In \Cref{tab:f1_auc_sota}, we summarize the effectiveness of \method{} and the other methods by showing the best F1 and PR-AUC scores. Results verify that \method{} is the most effective method for both join benchmarks across all similarity thresholds. This finding is of high importance, as the effectiveness of the other methods can fluctuate depending on the underlying datasets. 

\mybox{\textbf{Takeaways:} i) \method{} is consistently more effective than the state-of-the-art baselines, and ii) other methods exhibit low precision when the recall is high, while \method{} provides far fewer false join predictions.}

\begin{figure*}[t!]
    \centering
    \hspace{3mm}\begin{minipage}{\columnwidth}
            \includegraphics[width=.8\columnwidth]{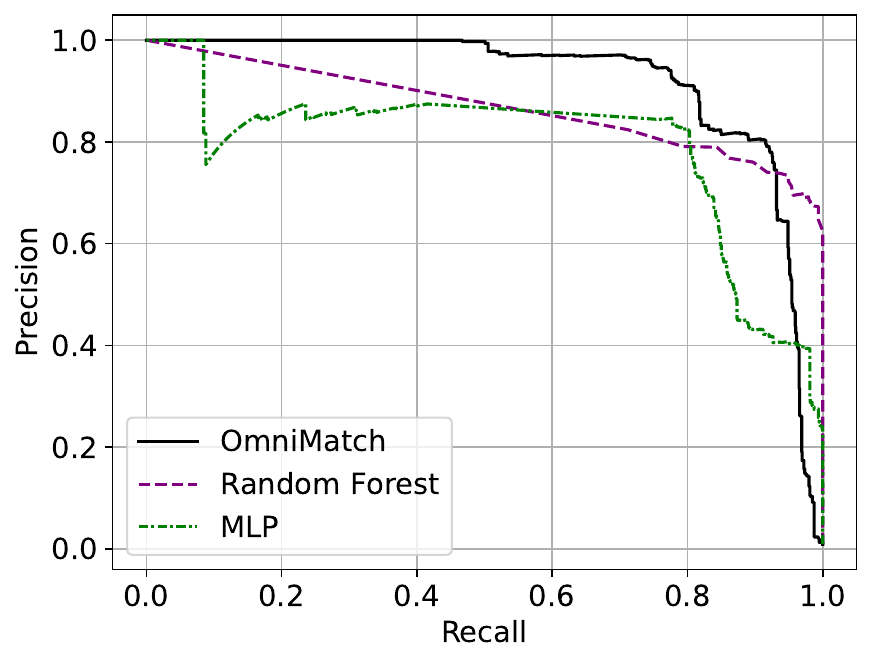}
            \vspace{-2mm}
            \subcaption{City Government}
            \label{fig:comparison_cg_ml}
    \end{minipage}
    \begin{minipage}{\columnwidth}
             \includegraphics[width=.8\columnwidth]{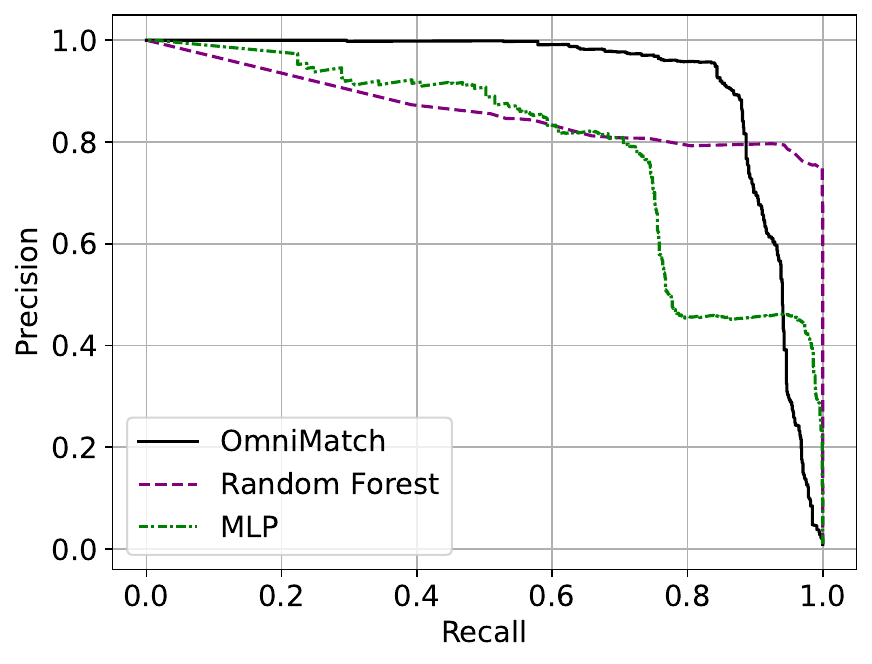}
             \vspace{-2mm}
             \subcaption{Culture Recreation}
             \label{fig:comparison_cr_ml}
    \end{minipage}
    \vspace{-2mm}
    \caption{Effectiveness comparison of \method{} with other ML-models for various similarity thresholds.}
    \label{fig:comparison_ml}
\end{figure*}

\subsection{Comparison to Other ML Models}

We assess the gains in effectiveness of \method's training model in comparison to other ML baselines that utilize only the column pairwise similarities. \autoref{fig:comparison_ml} shows the Precision-Recall of our method compared to the Random-Forest (RF) and MLP models for both join benchmarks. The main observation here is that, regardless of the underlying datasets, \method's join prediction model is superior to the other two, as it achieves considerably higher precision for the majority of recall values; the RF model achieves higher precision only when the similarity threshold is too low (thus, a threshold that would not be used in a realistic scenario). This result highlights the effectiveness of our graph modeling: \method's RGCN column representations better capture column join relationships and avoid false positive predictions of models that rely only on the column pairwise similarities.

Results in \autoref{tab:f1_auc_ml} verify \method's improvements in overall effectiveness as opposed to using less sophisticated ML models. Specifically, our method produces the highest overall F1-score, i.e., it can predict more accurate join relationships than pairwise similarities in conjunction with either an RF or MLP model. In addition, the high PR-AUC scores further showcase that \method{} consistently achieves high precision regardless of the similarity threshold used to decide whether a column pair represents a valid join. In contrast, using only the column pairwise similarities cannot help the RF and MLP models to capture less direct join relationships, while it can critically increase false-positive rates.

\mybox{\textbf{Takeaway:} \method's prediction model, using the column representations produced by the RGCN model, leverages column pairwise similarities to result in significantly better effectiveness than less sophisticated prediction models.}

\subsection{Ablation Study: Effect of Similarity Signals}
\begin{figure}[t]
    \centering
    \includegraphics[width=0.9\columnwidth]{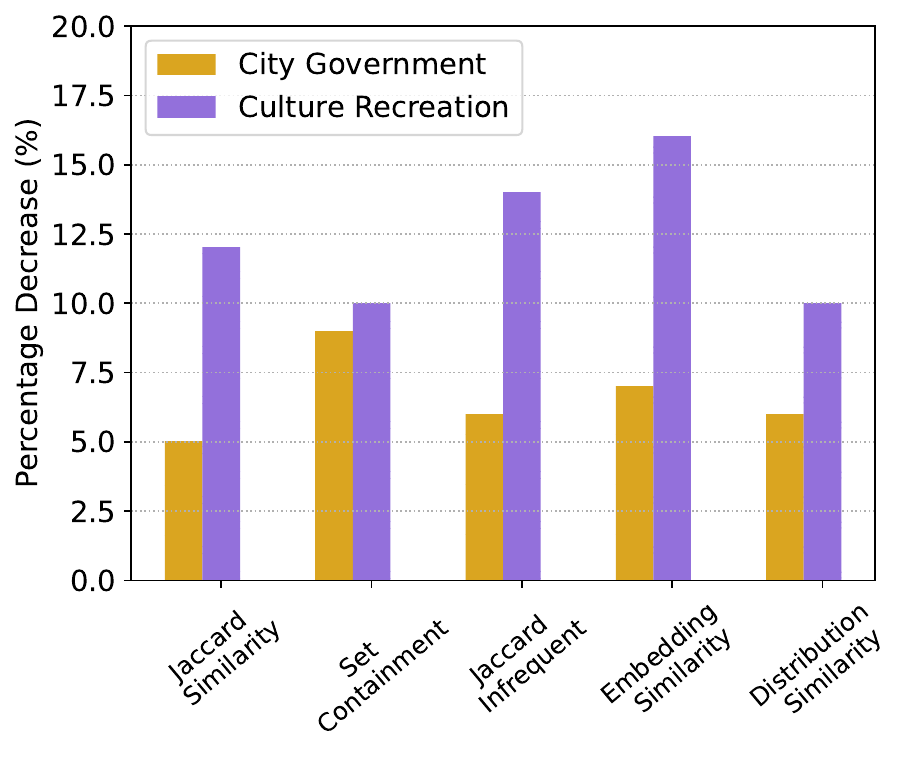}
    \caption{Reduction (in Percentage Decrease) of best F1-scores when \method{} considers a single similarity signal.}
    \label{fig:per_dec}

\end{figure}

\reviewer{2: W2- Third, as transitivity is considered effective in [19], it is necessary to show ablation on whether RGCN provides features other than simply applying graph transitive closure (first point in W1) to show that RGCN is a necessary and sound solution.}
We evaluate the power of using multiple similarity signals to construct our graph, in contrast to considering single ones. In \Cref{fig:per_dec}, we show the percentage decrease in the best F1-score achieved by \method{} when considering only one similarity signal per run. First, in \Cref{fig:per_dec}, we see that the results support our intuition: using only one signal to build the similarity graph considerably affects the ability of our model to decide correctly on whether a column pair represents a join. Indeed, relying on single similarities incurs drops in the best F1 scores achieved due to increasing false positive rates. Moreover, many valid column join cases in our benchmarks have yet to be discovered by \method{} when employing single similarity signals due to information loss of transitive paths in the constructed similarity graph. For instance, using only Jaccard similarity can severely harm the effectiveness of capturing fuzzy joins since it checks only for exact value overlaps.

\begin{table}[tb]
\centering
\resizebox{\columnwidth}{!}{
\begin{tabular}{l|c||c||c}
\toprule
\multicolumn{4}{c}{\textbf{Best F1 Scores}}\\
\bottomrule
\textbf{Benchmark} & \textbf{\method} & \textbf{Random Forest} & \textbf{MLP} \\ \toprule
\bottomrule
\textbf{City Government} & \textbf{0.857} & 0.827 & 0.813 \\ \hline
\textbf{Culture Recreation} & \textbf{0.894} & 0.862 &  0.755 \\
  \toprule
  \multicolumn{4}{c}{\textbf{PR-AUC Scores}}\\
\bottomrule
\textbf{Benchmark} & \textbf{\method} & \textbf{Random Forest} & \textbf{MLP} \\ \toprule
\bottomrule
\textbf{City Government} & \textbf{0.920} & 0.805 & 0.788\\ \hline
\textbf{Culture Recreation} & \textbf{0.921} & 0.835 & 0.618\\
  \bottomrule
\end{tabular}
}
\caption{Best F1 and PR-AUC scores comparison of \method{} \& other ML models.}
\vspace{-20pt}
\label{tab:f1_auc_ml}
\end{table}
In addition, a crucial observation here is that the percentage decrease vastly relies on the underlying datasets and column joins to be captured. As we see in \autoref{fig:per_dec}, the drop in best F1-scores is significantly higher on the Culture Recreation benchmark with percentage decrease values of at least \texttt{10$\%$}. This is due to the following two reasons: \emph{i}) there are column pairs in this benchmark that share (partial) value overlaps (e.g., dates), whereas they do not represent join relationships and \emph{ii}) most column joins in the City Government benchmark are more distinguishable, i.e., a potential (partial) value overlap strongly indicates a valid join.

No similarity signal consistently incurs larger/smaller effectiveness drops across the two join benchmarks. For instance, using only set containment leads to the lowest percentage decrease in the best F1 score for the Culture Recreation benchmark and the highest for the City Government one. This observation reinforces our claim that no similarity signal can be fully trusted when isolated from the rest since its effectiveness depends on the characteristics of the underlying datasets. Only the complete set of similarity signals used in \method{} can provide the best join discovery results.

\mybox{\textbf{Takeaways:} i) using a single similarity signal incurs a notable decrease in effectiveness, and ii) \method's consistency is strongly connected to using a comprehensive set of the proposed similarity signals.}

\subsection{OmniMatch Execution Times}
While we consider any-join discovery as an offline procedure in data repositories (as opposed to online procedures such as dataset search in  data lakes), for the sake of completeness we report in Table \ref{tab:times} the execution times of \method{} for both join benchmarks. Specifically, we report the time for the steps we show in Figure \ref{fig:training-inference}, \emph{i}) generating joinable pairs and transforming them into a similarity graph (T1 $+$ T2), \emph{ii}) training \method's join prediction model (T3), \emph{iii}) building the similarity graph based on the original datasets (I1), and \emph{iv}) using the trained model for inference on it (I2). As expected, we see that the main bottleneck of our method is the similarity graph construction both for training and inference: computing the set of similarity signals and the initial node features for all column pair combinations for different datasets entails numerous column pairwise operations; yet, accelerating these computations is a trivial issue that is not in the scope of this work (e.g., when multiple cores are available they can be parallelized). On the other hand, we see that training times in both benchmarks are relatively small, especially when we consider that training takes place on a CPU; notably, training of state-of-the-art column embedding methods \cite{fan2022semantics, deepjoin} requires access to a GPU. Finally, the discrepancies we observe between the two benchmarks are due to the different number of columns and complexities of values stored in them.
\begin{table}[t!]
\centering
\resizebox{\columnwidth}{!}{
\begin{tabular}{l|c|c|c|c||c}
\toprule\bottomrule
\textbf{Benchmark} & \textbf{T1 $+$ T2} & \textbf{T3} & \textbf{I1} &\textbf{I2}  &  \textbf{Total} \\ \toprule\bottomrule
\textbf{City Government} & 48.8 &  7 & 53.6 & 0.5 & 109.9 \\ \hline
\textbf{Culture Recreation} & 23 & 8.4 &  8.7 & 0.5 & 40.6 \\
\toprule\bottomrule
\end{tabular}
}
\caption{\method{} execution times in minutes (CPU). T1-T3 and I1,I2 represent different steps of our method as shown in Figure 
\vspace{-20pt}
\ref{fig:training-inference}.}
\label{tab:times}
\end{table}

%% file: 7_Related_Work.tex
\section{Related Work}
\label{sec:related}

\asterios{to shorten this}

We have already gone through essential works in \Cref{sec:introduction} and \Cref{sec:exp}. In this section, we discuss related work relevant to join discovery, including schema matching (\Cref{sec:rel-matching}) and dataset search/discovery (\Cref{sec:related:search}).

\subsection{Schema Matching} 
\label{sec:rel-matching}

\para{Traditional Matching Methods} Schema matching on tabular data includes automated methods for capturing relevance between columns of dataset pairs \cite{rahm2001survey}. These methods are mainly categorized into four categories: \emph{i}) \emph{schema-based} matching methods \cite{madhavan2001generic,melnik2002similarity} take into consideration only metadata at the schema level, such as column names, types etc., \emph{ii}) \emph{instance-based} methods that rely on the instances stored in the datasets to capture similarity among their columns, using signals like distribution similarity \cite{zhang2011automatic}, value overlaps and patterns \cite{engmann2007instance}, \emph{iii}) \emph{hybrid} ones that incorporated schema with instance information to predict column matches \cite{do2002coma}, and \emph{iv}) \emph{usage-based} methods that rely on query logs to build relatedness graphs among columns of datasets~\cite{elmeleegy2008usage}. 

\method{} is a self-supervised, instance-based method that can be used for any-join discovery. Contrary to other schema matching methods, \method{} is the first one to create column representations with RGCNs making use of multiple similarity metrics, outperforming COMA by 14\% on average (\Cref{sec:exp}). 


\para{Embedding-based Matching} Multiple embedding-based column matching methods have emerged, applying widely used methods for producing \textit{word embeddings} to encode table columns into the vector space and then to identify related columns in that space (e.g., with vector cosine similarity). To this end, pre-trained models such as Word2Vec \cite{mikolov2013distributed} and FastText \cite{bojanowski2017enriching}, have been applied to embed either column names \cite{fernandez2018seeping} or cell-values \cite{nargesian2018table}. In addition, \emph{locally-trained} embedding methods \cite{fernandez2019termite,koutras2020rema, cappuzzo2020creating} leverage the architecture of \textit{skip-gram models} \cite{mikolov2013distributed, bojanowski2017enriching} used in NLP, with extra pre-processing steps. Despite the seamless employment of methods using pre-trained models \cite{fernandez2018seeping, nargesian2018table}, or locally-trained embedding methods \cite{fernandez2019termite,koutras2020rema, cappuzzo2020creating}, they still seem to be insufficiently effective when used for matching related columns \cite{koutras2021valentine}. The latest and state-of-the-art embeddings-based methods Starmie~\cite{fan2022semantics} and DeepJoin~\cite{deepjoin}, use  Large Language Models (LLMs) and fine-tune them to create column embeddings for the needs of dataset discovery (unionable and joinable respectively).

Complementary to these methods, \method{} can use pairwise similarity metrics extracted from embedding-based methods (e.g., FastText value embeddings on infrequent tokens as discussed in Section \ref{ssec:similarities}). In our evaluation, we showcase our method's superior performance (14\% higher F1 and PR-AUC scores) with respect to state-of-the-art column embeddings methods \cite{fan2022semantics, deepjoin}.

\subsection{Related Dataset Search/Discovery}
\label{sec:related:search}
Given a dataset $Q$ as a query, dataset search methods focus on returning the top-$k$ related datasets for $Q$. Relatedness refers to either table \emph{unionability} \cite{nargesian2018table, bogatu2020dataset, fan2022semantics, khatiwada2022santos} or table \emph{joinability} \cite{fernandez2018aurum, zhu2019josie, dong2021efficient}. Typically, related-dataset search methods use column similarity signals (as used in schema matching methods \cite{koutras2021valentine}) between column pairs to generalize relatedness scores between datasets. Contrary to top-$k$ dataset search that focuses on returning the top-$k$ dataset, given a query dataset, 
\method{} focuses on the problem of column joinability discovery (returns pairs of joinable columns, not datasets). At the same time, \method{} draws inspiration from pairwise column similarities that have been used in the related dataset search literature (Section \ref{ssec:similarities}). In addition, we have adapted column embeddings from the state-of-the-art dataset discovery methods Starmie \cite{fan2022semantics} and DeepJoin \cite{deepjoin}, for the needs of returning joinable column pairs as described in \Cref{sec:exp}, where \method{} outperforms them by 14\% in F1 and PR-AUC scores.

%% file: 8_Conclusion.tex
\section{Conclusion}
\label{sec:conclusion}

In this paper, we introduced \method{}, a novel self-supervised method that captures joins of any kind across tabular data of a given repository. \method{} leverages a comprehensive set of similarity signals and the transitive power of a graph model to learn column representations based on an RGCN. Notably, our method can automatically generate positive and negative join examples to guide the learning process. Our experimental evaluation shows that \method{} is considerably more effective than state-of-the-art column matching and representation methods. In contrast, our prediction model based on RGCNs is substantially more accurate than others. In addition, we justify the gains of using the comprehensive set of similarity signals we propose.
